# X-Ray-Driven Photon Bunching

**Shaul Katznelson[1†], Noam Kasten[1†], Offek Tziperman[1], Avner Shultzman[1], Tomer Bucher[1], Tom Lenkiewicz Abudi[1], Roman Schuetz[1], Orr Be'er[2], Shai Levy[2], Rotem Strassberg[1], Georgy Dosovitsky[2], Sotatsu Yanagimoto[4], Francis Loignon-Houle[5], Yehonadav Bekenstein[1,2], Charles Roques-Carmes[3], and Ido Kaminer[1]**

1 Solid State Institute, Technion – Israel Institute of Technology, Haifa 3200003, Israel

2 Faculty of Materials Science and Engineering, Technion – Israel Institute of Technology, Haifa 3200003, Israel

3 E. L. Ginzton Laboratory, Stanford University, Stanford, California 94305, USA

4 Department of Materials Science and Engineering, Institute of Science Tokyo, Midoriku, Yokohama, 226-8503, Japan

5 Instituto de Instrumentación para Imagen Molecular (I3M), CSIC — Universitat Politècnica de València, Valencia, Spain

† Equal contribution

## Abstract

Hanbury Brown and Twiss (HBT) interferometry is a milestone experiment that transformed our understanding of the nature of light. The concept was demonstrated in 1956 to measure the radii of stars through photon coincidence detection. This form of coincidence detection later became a cornerstone of modern quantum optics. Here we connect HBT interferometry to the physics of scintillation, the process of spontaneous light emission upon excitation by high-energy particles, such as x-rays. Our work reveals intrinsic photon bunching in the scintillation process, which we utilize to elucidate its underlying light emission mechanisms. Specifically, $g^{(2)}(\tau)$ enables the quantitative extraction of scintillation lifetime and light yield, showing their dependence on temperature and X-ray flux as well. This approach provides a characterization method that we benchmark on a wide gamut of scintillators, including rare-earth-doped garnets and perovskite nanocrystals. Our method is particularly important for nano- and micro-scale scintillators, whose properties are challenging to quantify by conventional means: We extract the scintillation properties in perovskite nanocrystals of only a few hundreds of nanometers, observing strong photon bunching ($g^{(2)}(0) > 50$). Our research paves the way for broader use of photon-coincidence measurement and methods from quantum optics in studying materials with complex optical properties in extremes regions of the electromagnetic spectrum.

**Introduction**

Scintillation, the process converting high-energy radiation quanta into UV-to-visible light, has been a central pillar of research and applications of high-energy particles and ionizing radiation since the late 19th century [1]. The main use of scintillators is in detection of such radiation, making scintillators a vital part of everyday technologies and of fundamental research efforts [2]. Scintillators play a crucial role in medical imaging [3], security scanners [4], non-destructive testing [5], and several fields of scientific research including cosmology, nuclear, and high energy physics [6]. Ongoing research efforts strive to improve scintillation technology: for example, faster scintillators can improve the resolution of medical imaging based on time-of-flight procedures [7].

In recent years, new classes of scintillators have emerged, leveraging novel nanomaterials such as quantum dots [8,9] and photonic crystals [10]. Such scintillators promise faster light emission with higher yield, for both new and existing applications [11,12]. Despite the proliferation of efforts in creating these new scintillators, a challenge persists in the characterization of their intrinsic properties. This challenge is especially acute in nanomaterials [13–18], which rely on low-dimensional materials to enhance the scintillation process. Such extremely thin geometries are challenging to study, because conventional characterization methods are typically tailored for thick bulk scintillators [6,19].

The underlying physics of ionizing radiation to light conversion in scintillators is a complex and multi-step process: particle collisions and energy loss, excitation of hot carriers and their dynamics, electron-hole pair diffusion, and spontaneous emission [2]. The initial collision of high-energy photons (Figure 1a) creates hot carriers propagating through the material (Figure 1b), initiating complex dynamics and generating many electron-hole pairs that can recombine, subsequently creating visible photons via spontaneous emission (Figure 1c). This complex process, in both conventional and novel scintillators, motivates new types of experiments to better understand and use its underlying physics.

Here we find that the intrinsic characteristics of the scintillation process are imprinted on the photon statistics of its light emission. We observe that light emission in scintillation is intrinsically bunched – a universal feature that is shared by all scintillator materials that we benchmarked with this method. Further analysis of the bunching statistics reveals the key parameters of the internal scintillation process of each material. Specifically, we measure the second-order photon correlation $g^{(2)}(\tau)$ using Hanbury-Brown-Twiss (HBT) interferometry [20,21] and demonstrate how to extract from it the light yield and emission lifetime of scintillation (Figures 1c,d). Leveraging this characterization method, we discover strong bunching in perovskites nanocrystals. Our method opens the way to deeper understanding of complex processes in advanced materials by relying on measurements of photon statistics by coincidence methods.

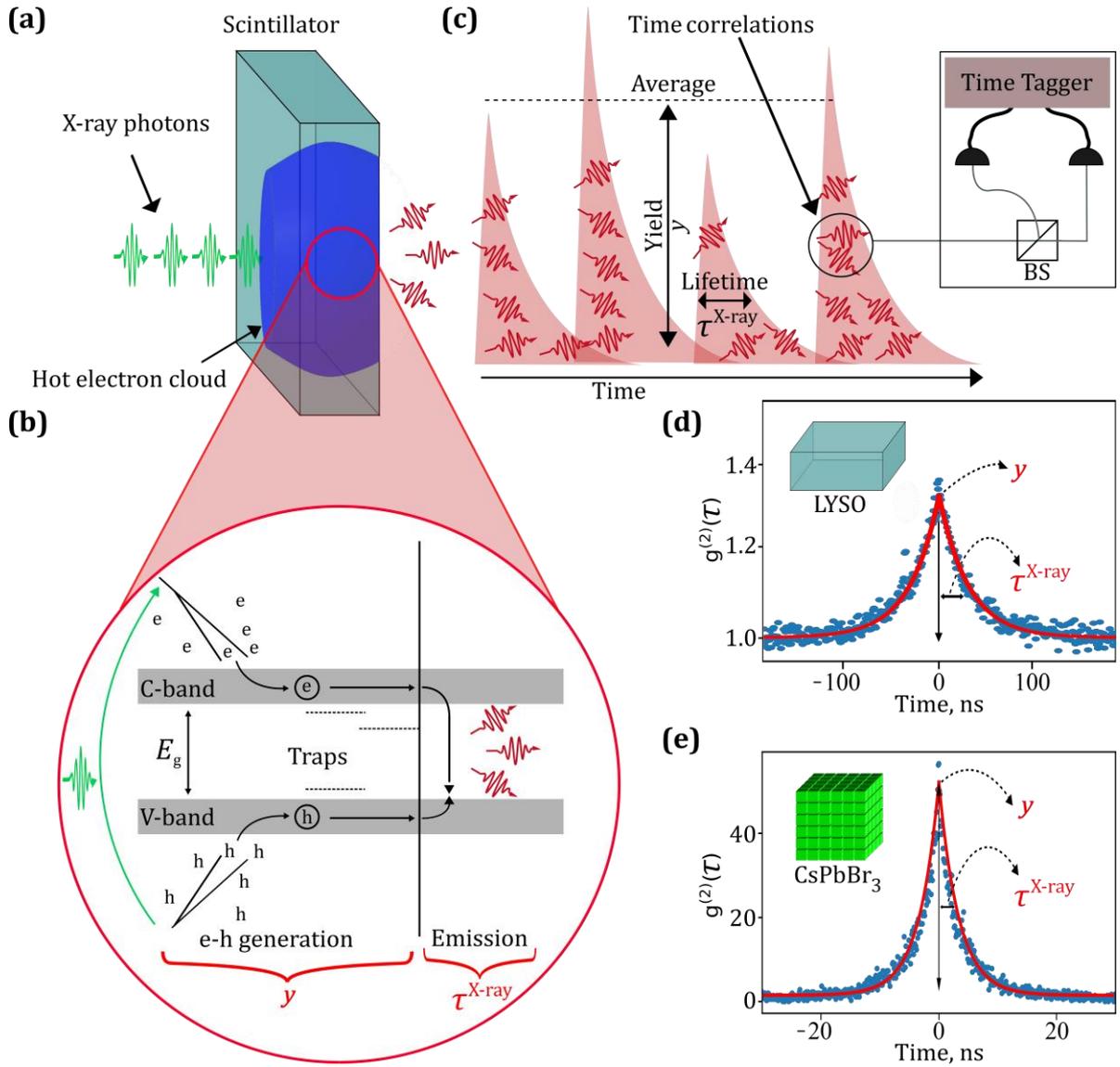

*Figure 1: Schematic of scintillation process and parameter extraction. (a) Scintillator irradiated by X-rays (cyan waves) creating an electron cloud within the scintillator (blue shape). (b) A simplified schematic of the scintillation process within the scintillator. (c) Illustration of the light emission and of two scintillation parameters that we can extract using the HBT measurement: light yield (y) and scintillation lifetime ($\tau^{X-ray}$). The inset illustrates the collection of two time-correlated photons. (d,e) Second-order correlation $g^{(2)}(\tau)$ measurements of LYSO:Ce single crystal (d) and CsPbBr₃ nanocrystal superlattices (e), with the raw data marked by blue dots, and its exponential fit by a red curve. The insets depict the samples. These measurements allow us to extract the scintillation parameters for bulk and nano scintillators alike. X-ray excitation parameters are detailed in section S1.*

**Photon correlations in scintillation**

The scintillation process (Figure 1b) is described by a few key parameters: light yield ($y$) and lifetime ($\tau^{X-ray}$). The lifetime depends on electron-hole recombination and spontaneous emission. The light yield also depends on hot-carriers dynamics, electron-hole generation, and their transport properties. These parameters are crucial for most applications of scintillators across a wide range of fields. Notable examples are time-of-flight (TOF) and coincidence time resolution (CTR) in positron emission tomography (PET) [7], which utilizes the coincidence between pairs of γ photons arriving at detectors positioned opposite to each other to extract the position of an annihilation event in the human body. The coincidence time resolution in this medical imaging modality is directly influenced by the scintillator intrinsic properties. Consequently, accurate estimation of the light yield and lifetime is essential for the design of new scintillators with desirable characteristics.

We developed a characterization method for scintillator materials that relies on photon correlation measurements with HBT interferometry, which was originally developed for measuring the photon fluctuations and correlations of light [20]. The HBT setup consists of a 50/50 beam splitter and two single-photon detectors (Figure 1c inset). Measuring the time correlation between detection events, it is possible to determine the second-order coherence/moment of the photon statistics of the measured light. This measurement is denoted by $g^{(2)}$.

HBT photon-correlation detection classifies light sources into three regimes: (1) Poissonian distributed light as in conventional emitters (e.g., LEDs) [22] for which $g^{(2)}(0) = 1$; (2) Sub-Poissonian or anti-bunched light (e.g., single photon emitters) [23] for which $g^{(2)}(0) < 1$; (3) super-Poissonian or bunched light (e.g., squeezed or thermal light) [24,25] for which $g^{(2)}(0) > 1$.

Our work shows not only that scintillation light is bunched ($g^{(2)}(0) > 1$), but also how this photon statistics enables us to extract the intrinsic properties of the scintillation process (Figure 1d,e). The bunching arises from the nature of the scintillation process, whereby each individual X-ray photon initiates an avalanche process resulting in emission of multiple photons in the UV-to-visible range. Our methodology exemplifies a broader concept of measuring photon statistics and correlations to extract hard-to-access properties of optical materials.

**Extracting scintillation lifetime from photon correlations**

Our experimental concept relies on HBT to extract the scintillation light yield ($y$) and lifetime ($\tau^{X-ray}$) for a wide range of scintillator materials. Two example HBT $g^{(2)}(\tau)$ results are shown in Figure 1d,e, for a standard Cerium-doped Lutetium–Yttrium Orthosilicate (LYSO:Ce) scintillator and for a <1μm scintillator film based on perovskite quantum-dot superlattices. The latter result emphasizes that our approach can directly characterize otherwise hard-to-measure nanometric thin scintillators (see S4 for further analysis).

Figure 2a presents our measurement setup, highlighting that a conventional continuous-emission X-ray tube is sufficient to characterize the scintillators lifetime and other properties. The X-rays impinge a scintillator sample, emitting visible light that is then focused via a microscope objective lens (OBJ) into a fiber. After a 50/50 fiber beam splitter (BS), the light is detected by two single photon avalanche detectors (SPADs), whose signal is analyzed using a time tagger. See additional experimental details in S1.

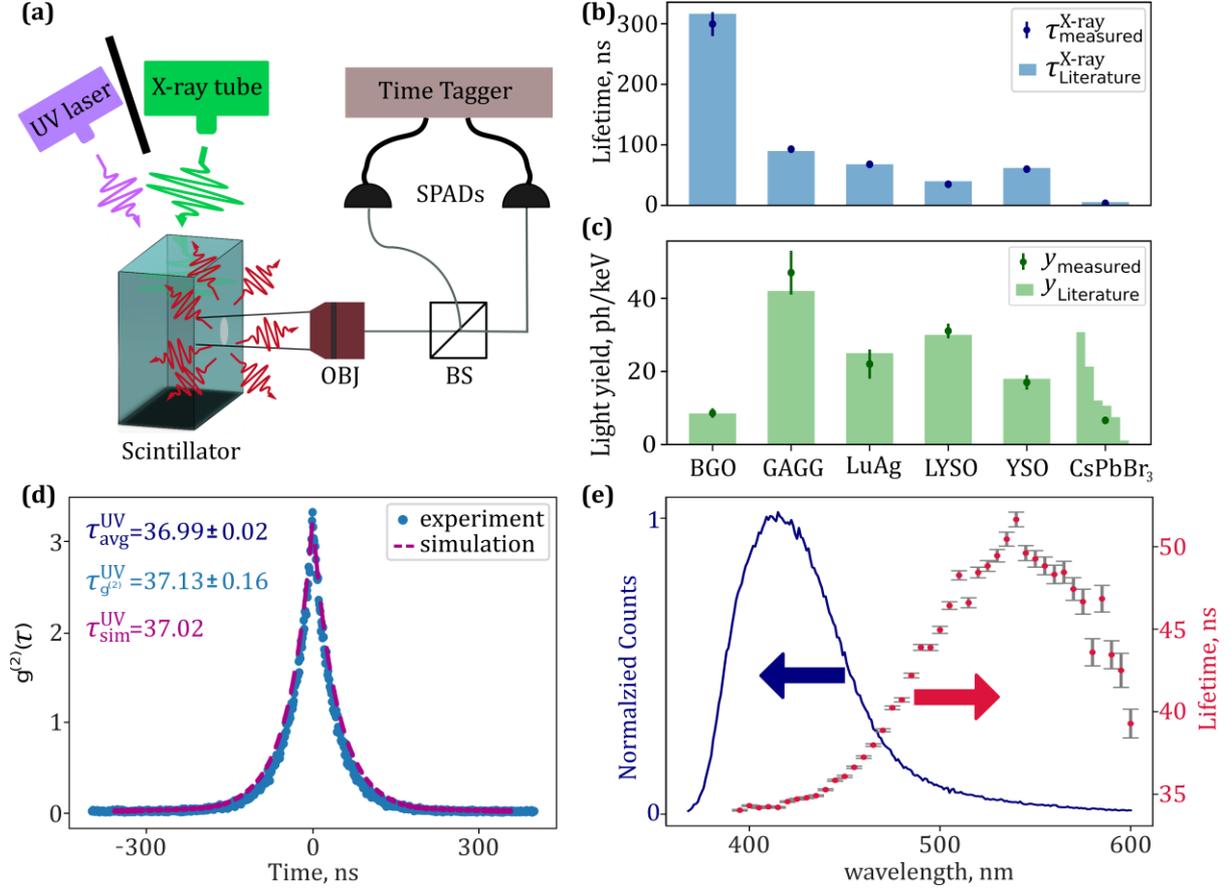

*Figure 2: Investigation and extraction of Scintillator lifetime and light yield. (a) Schematic of the experimental setup. A scintillator is excited by either X-rays (green) or by a UV laser (blue), emitting visible photons (red waves). These photons are then collected with an objective lens (OBJ), collimated into a 50/50 beam splitter (BS), and detected by two Single Photon Avalanche Detectors (SPADs), correlated using a Time Tagger. (b-c) Scintillation lifetime and light yield, comparing the measured values (dots) with the manufacturer specifications [26] (horizontal bars) for the conventional scintillators and with literature [27–29] for $CsPbBr_3$. (d) To calibrate our method, we perform analogous $g^{(2)}(\tau)$ measurements of LYSO (blue) under UV excitation (purple waves in (a)), compared with a simulation (cyan) using direct lifetime measurements. (e) Direct lifetime measurements for different emission wavelengths (red) and emission spectrum (blue) of LYSO under UV excitation.*

We extract the scintillation lifetime from an exponential fit to the line shape of $g^{(2)}(\tau)$, and calculate the light yield from the integral on $g^{(2)}(\tau)$ combined with the total count rate $\Gamma_0$ (Figure 2b,c). In order to calibrate and verify that the line shape of $g^{(2)}(\tau)$ provides the lifetime as expected, we conducted a comparative analysis using HBT under UV illumination (Figure 2d) and using a direct lifetime measurement based on a pulsed light-emitting diode-UV laser (Figure 2e). The HBT-based lifetime $\tau_{g^{(2)}(\tau)}^{UV} = 37.13 \pm 0.16$ ns is consistent with the weighted average lifetime based on direct measurements $\tau_{avg}^{UV} = 37.26 \pm 0.78$ ns. For a full theoretical description and additional support of the method, we also conducted a Monte Carlo simulation (see S4) based on the same measurements and extracted $g^{(2)}(\tau)$ from this simulation, yielding $\tau_{sim}^{UV} = 37.02 \pm 0.21$ ns]. These values are also in agreement with the manufacturer specification [26].

Figures 2b,c summarize extracted lifetimes and light yields for several scintillators, showing a good match of our measurements to data from the literature. Specifically, we tested Bismuth Germanate (BGO,) as well as Cerium-doped Gadolinium Aluminium Gallium Garnet (GAGG), Lutetium Aluminium Garnet (LuAG), LYSO and Yttrium Orthosilicate (YSO) [26]. This match adds confidence in our method and supports applying it for novel scintillators such as lead-halide perovskite nanocrystals embedded in a polystyrene matrix (right-most column in figures 2d,e). Measurements of scintillation light yield and lifetime for such materials are challenging, especially for nanocrystals like the ones we measured. Reports so far varied drastically, between 0.3-16 ph/keV, with a few papers reporting over 20 ph/keV [27–29].

**Theoretical framework and match to experiments for precise quantification**

To support the experimental design and scintillators characterization, we develop a theoretical framework linking $g^{(2)}(\tau)$ with the scintillation process and its physical characteristic parameters. This framework is inspired by works on cathodoluminescence, first by Meuret et al. [30], and later expanded to a range of materials [31,32] and to more general theoretical frameworks [33]. Building on these works, we develop a theory capturing the complete dependence of $g^{(2)}(\tau)$ on the intrinsic scintillator properties, enabling to extract the scintillation lifetime ($\tau^{\text{X-ray}}$), light yield ($y$), and absorption efficient ($\eta_{\text{sample}}$). The full analytical derivation is presented in S4.

The following two equations summarize the implications of our theory under two simplifying assumptions: the absorption is dominated by the photoelectric effect (considering the low-energy X-ray photons and the high-Z materials [34]) and the number of excited emitters per X-ray photon has a low-variance distribution.

The photon bunching value for scintillators is given by:

$$g^{(2)}(\tau) = 1 + \frac{1}{2\tau^{\text{X-ray}} \cdot J \cdot \eta} \cdot e^{-\frac{|\tau|}{\tau^{\text{X-ray}}}}, \quad (1)$$

where $J$ denotes the input flux of the high-energy particles (X-ray photons in our case) and $\eta = \eta_{\text{sample}} \cdot \eta_{\text{system}}$ is the overall X-ray excitation efficiency. The term $\eta_{\text{sample}}$ refers to the scintillator's absorption efficient, which is determined by the material properties and scintillator geometry. The term $\eta_{\text{system}}$ is defined by the X-ray system efficiency, which accounts for losses from the optical setup, is attributed to coupling losses and finite generation efficiency.

To bring additional experimental evidence of the relation between $g^{(2)}(\tau)$ and $\tau^{\text{X-ray}}$, we measured two scintillators at temperatures ranging from 77-300K. For each scintillator, the lifetime changes in opposite direction at lower temperatures, yet the inverse lifetime scaling of equation (1) remains valid. This analysis is detailed in S3.

To extract the scintillator parameters from a $g^{(2)}(\tau)$ measurement, we have developed a novel method to combine this photon correlation data with the total count rate $\Gamma_0$. We calibrate for the system loss factor, eliminating the absorption efficiency and the detection efficiency, by using a reference scintillator (detailed in section S4), yielding:

$$y_B = y_A \cdot \frac{Q_A}{Q_B} \cdot \frac{\Gamma_{0B}}{\Gamma_{0A}} \cdot \frac{\left(g_B^{(2)}(\tau) - 1\right)_{Area}}{\left(g_A^{(2)}(\tau) - 1\right)_{Area}} \quad (2)$$

Here, A denotes the reference scintillator, B denotes the scintillator we wish to study and $\left(g^{(2)}(\tau) - 1\right)_{Area}$ denotes the area under the curve of the specified $g^{(2)}(\tau) - 1$ function. Using equations (1-2), we characterize several scintillators and show that this method is robust to arbitrary thicknesses and material properties (Figure 2d,e and further details in section S2). In contrast, standard methods relying solely on $\Gamma_0$ as described in section S7.3 exhibit high variance, highlighting the importance of incorporating the $g^{(2)}(\tau)$.

Using this framework, we also investigate the dependence of the photon bunching on the X-ray flux $J$ (Figure 3). We conducted a comparative analysis on three scintillator materials subjected to varying X-ray flux levels controlled by the X-ray tube current. Figure 3 (a-c) presents the scintillation spectrum of BGO, LYSO(Ce), and CsPbBr3 perovskites nanocrystals in a polystyrene matrix, left to right respectively. Higher X-ray flux increases the overall signal for each material, enabling calibration of the X-ray flux for each current value. Figure 3 (d-f) compares the HBT $g^{(2)}(\tau)$ for different materials under identical flux conditions. As predicted by equation (1), the height of $g^{(2)}(0)$ is inversely proportional to the X-ray flux (Figure 3 (g-i)), analogous to findings from studies in cathodoluminescence [30].

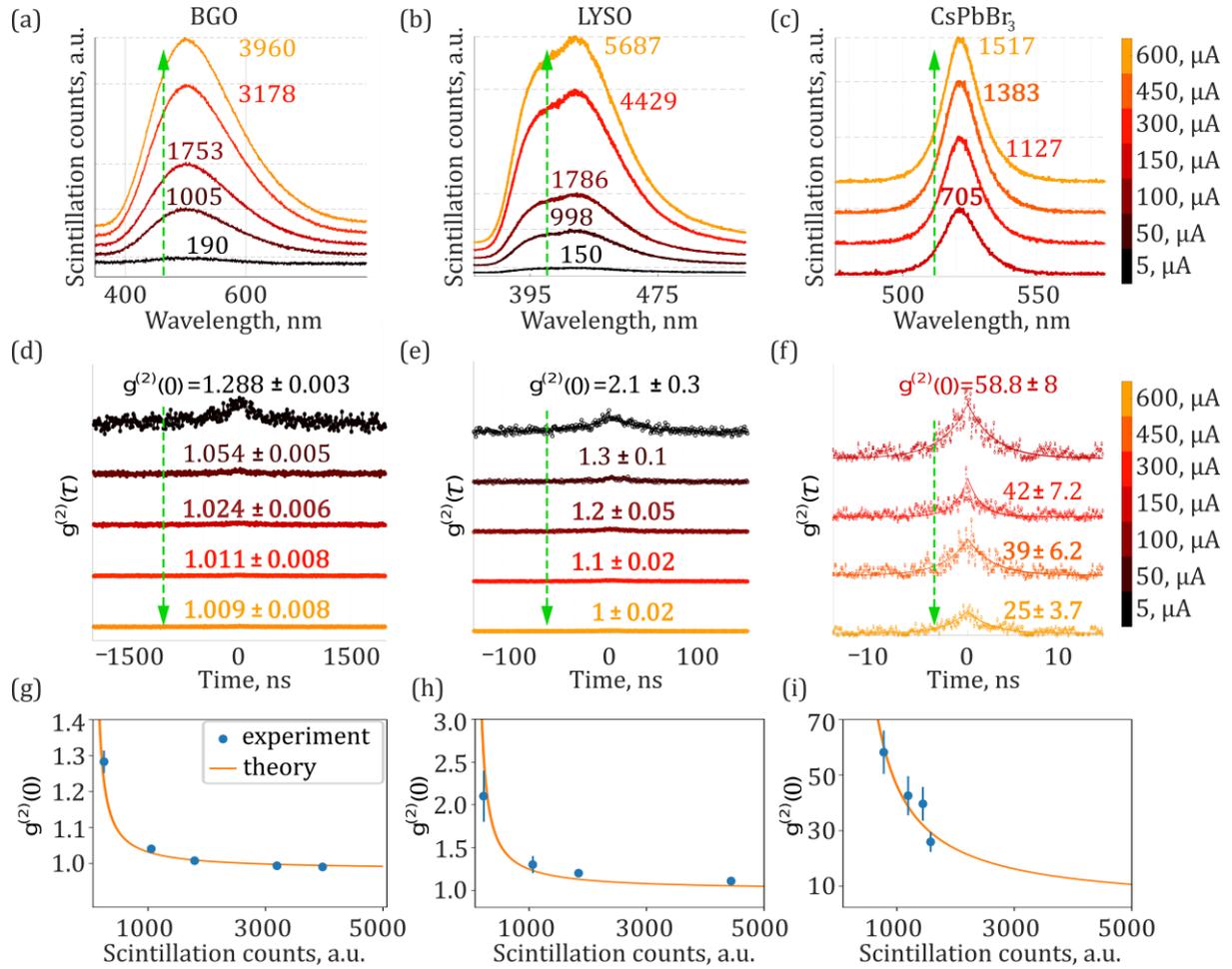

*Figure 3: Flux-dependence of photon bunching in scintillation*, shown for BGO, LYSO, and CsPbBr$_3$ nanocrystals, in the left, middle, and right columns respectively. *(a-c)* Spectral response for different values of current operating the X-ray tube. The numbers represent the maximum counts for each curve. *(d-f)* The HBT $g^{(2)}(\tau)$ for different values of current operating the X-ray tube. The numbers indicating the value of $g^{(2)}(0)$ for each curve. *(g-i)* The relation between the maximal spectral response and $g^{(2)}(0)$.

## Discussion and outlook

Our HBT approach for scintillation characterization complements conventional methods in the field. One common approach relies on a pulsed X-ray source with a streak camera or another ultrafast optical detector [31]. Other techniques utilize X-ray and gamma-ray pulses generated in large facilities such as synchrotrons for pump-probe characterization of the scintillators [35]. Another commonly used technique employs radioactive sources such as $^{137}$Cs or $^{22}$Na to produce pairs of high-energy gamma photons, which can be detected using a start-stop method with two detectors [31,32,35]. Alternatively, these gamma photons can be coupled to a single photomultiplier tube and oscilloscope to measure both the lifetime and light yield [36]. Furthermore, different radioactive sources may be used to produce alpha particles, which have shorter absorption lengths compared to X-rays, allowing for efficient interaction with thin layers and powdered scintillators [37]. Our approach offers practical advantages in terms of simplicity, robustness, and high sensitivity, particularly when detecting low-energy X-ray excitations in dim scintillators.

In conclusion, inspired by quantum optical analysis of light emission, we observed photon bunching in scintillators and showed what insight it provides about the scintillation process. Through a quantitative study, we propose a HBT photon-correlation setup for scintillator characterization, showing its particular advantage for hard-to-characterize thin scintillators from novel nanomaterials.

Further developments of our work can lead to *enhanced X-ray microscopy* that relies on photon correlations. For example, we implemented a raster scan over a $CsPbBr_3$ scintillator for which we extract the $g^{(2)}(0)$ value at each point, reconstructing a two-dimensional correlation image (see S5). While not yet optimized for enhanced microscopy, this raster scan demonstrates the possibility of integrating a scanning-microscopy technique with our photon-correlation scintillation setup.

Looking forward, non-trivial emission statistics may be used in various correlation-based imaging methods, such as ghost imaging [35,38–40]. We envision photon-correlation measurements of $g^{(2)}$ and higher orders to serve as a novel approach to measuring the intrinsic properties of complex optical materials, such as nonlinear-optical media dominated by charge transport phenomena and by hot carrier dynamics. We anticipate a continued rise in adoption of quantum optical methods and experimental apparatus for applications in microscopy and spectroscopy of various optical materials, especially ones governed by complex condensed-matter processes.